# Centralized Lifetime Maximizing Tree For Wireless Sensor Networks

First Deepali Virmani , Second  Satbir Jain

*ABSTRACT -* To enable data aggregation among the event sources in wireless sensor networks and to reduce the communication cost there is a need to establish a coveraged tree structure inside any given event region to  allow data reports to be aggregated at a single processing  point prior to transmission to the network. In this paper we propose a novel technique to create one such tree which maximizes the lifetime of the event sources while they are constantly transmitting for data aggregation.   We use the term Centralized Lifetime Maximizing Tree (CLMT) to denote this tree. CLMT features with identification of bottleneck node among the given set of nodes. This node collects the data from every other node via routes with the highest branch energy subject to condition loop is not created. By constructing tree in such a way ,protocol is able to reduce the frequency of tree reconstruction, minimize the delay and maximize the functional lifetime of source nodes by minimizing the additional energy involved in tree reconstruction.

## I. INTRODUCTION

Wireless Sensor Networks (WSNs) may deploy several hundreds to thousands of sensor nodes. Protocols in such networks must therefore be scalable .Unlike the conventional ad hoc communication networks, energy resources in WSNs are usually scarce due to the cost and size constraints of sensor nodes. In addition, it is impractical to replenish energy by replacing batteries on these nodes. Conserving energy is thus the key to the design of an efficient WSN. Most sensor nodes are task-specific in that they are all programmed for one common application. A node at one specific time may be granted more access to the network than all other nodes if the program objective is still satisfied. For this reason, network resources are shared but it is not necessary that they may be equally distributed as long as the application performance is not degraded Since sensors are being densely deployed in WSNs, the detection of a particular stimulus can trigger the response from many nearby nodes. Thus, data in such networks are usually not directly transmitted to interested users upon event detection. Instead, they are aggregated with neighboring sources locally to remove any redundancy and produce a more concrete reading [1], [2], [3], [4]. In this paper, we focus on constructing a data aggregation tree among any given set of source nodes. The tree has a dedicated root for which the data from various sources are gathered. Moreover, the tree is structured in a way that can preserve the functional lifetime of the event sources subject to the condition that they are constantly transmitting. The functional lifetime is defined as the time till a node runs out of its energy. Reference [5] suggests that extending the node lifetime is equivalent to increasing the amount of information gathered by the tree root when the data rate is not time-varying. To shorten the time and minimize the energy cost to tree reconstructions, and hence maximizes the functional lifetime of all sources, we have proposed a Centralized Lifetime Maximizing Tree construction algorithm which arranges all nodes in a way to select a bottleneck node to collect the data in a centralized manner . Such arrangement extends the time to refresh the tree and lowers the amount of data loss due to a broken tree link before the tree reconstructions and maximizes the lifetime of the nodes. In fact, not all the trees are ideal for data collection inside the event region. Since energy is usually scarce in WSNs, it is most power-efficient if these sources can provide data to the sinks for the longest possible time. A tree that can survive for longer duration thus naturally becomes the best choice.

## II. Problem Formulation

Given a number of I connected source nodes with each source labeling i (i ε { 1,2 ……N }) and the knowledge of their own residual energy, $e_i$, our goal is to find a tree spanning all these sources and an appropriate tree root for data collection so that the functional lifetime of each source is maximized as much as possible. The time till the first link breaks in a given tree structure determines the lifetime of each source, and the term tree energy directly reflects this time [7],[8]. We hence tackle this problem by searching a tree that comprises the highest tree energy. The time during which data from each source along this branch can arrive at the root will depend on the minimum energy of any parent along this branch. By using the same analogy, the time during which data from all sources can arrive at the root without having to concern about broken link repairs and tree reconstructions will depend on the minimum energy of any branch, or equivalently that of any parent, in a given tree. The only question we are left with lies on how to select an appropriate tree root and the branch leading to each other source, such that the tree energy is maximized (Tree energy is defined as the minimum branch energy of all the branches in a given tree) . Before starting to describe our CLMT algorithm, we outline the basic spanning tree protocol [6] followed by presenting an energy-aware variant of it, namely E-Span [6].

- The conventional spanning tree fails to consider residual energy of nodes in the tree constructions. There is thus a good possibility that a low-energy node is arranged to forward data for some other nodes.

- E-Span[6] improves the design of tree construction by assigning root to be the highest energy node. Such arrangement provides root with the maximum amount of energy resources for its additional duty in coordinating the route to distant sinks. However, there is still a high chance of assigning low-energy nodes to be the data aggregating agents for the other sources

We start explaining the problem with E-Span[6] . The conventional spanning tree, E-Span is a cycle free graph which spans all the nodes as vertices. All other nodes are connected to the selected root via the shortest path route. Since the root , besides collecting data, is also responsible to coordinate the routes with distant sinks, so the node with the highest energy level is now chosen as the root. Moreover, each other node is given with the choice to select its parent as the highest-energy neighbor for which the shortest-path message comes from.. Unfortunately without having the complete knowledge of connectivity set provided by all sources some nodes in E-Span still traverse to the root through routes with lower branch energy. As a result each source is more often involved in tree reconstruction and utilizing a greater portion of its available energy in repairing the broken tree links over the course of its lifetime. Available functional lifetime of such nodes gets shorter due to additional energy cost involved in tree reconstruction. This is the major concern area with sensor networks . As the functional lifetime is the time till the first or set of nodes runs out of its energy [7], [8], [9], [10] or till the first loss of connectivity or coverage [11], [12] or a combination of these [13]. We aim at maximizing this functional lifetime. So we propose a novel technique to create a tree structure to maximize the lifetime using a centralized approach. We term this tree as Centralized Lifetime maximizing Tree ( CLMT) .

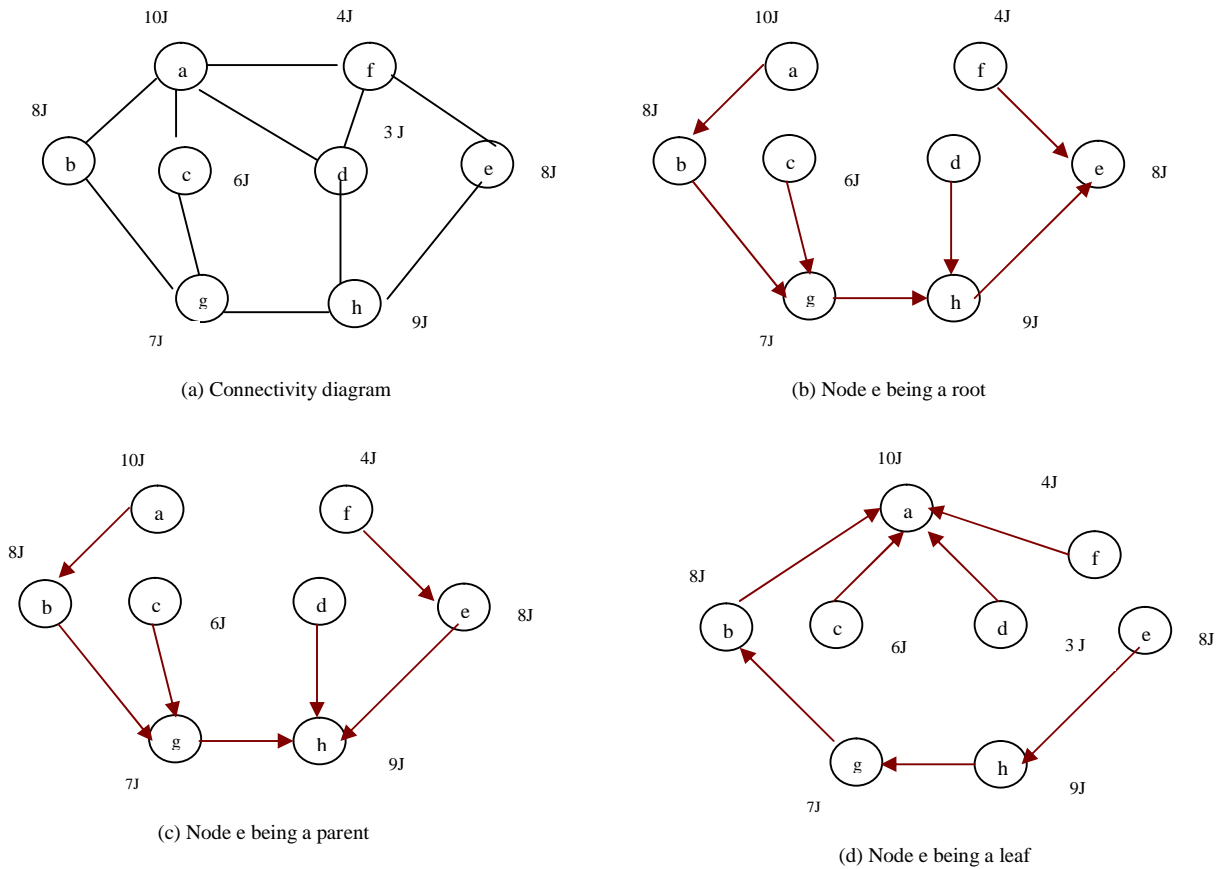

(a) Connectivity diagram

(b) Node e being a root

(c) Node e being a parent

(d) Node e being a leaf

**Figure 1  An example of bottleneck node**

## III. CENTRALIZED LIFETIME MAXIMIZING TREE

To minimize the energy cost to tree reconstructions, and hence preserve the functional lifetime of all sources, we have proposed a Centralized Lifetime maximizing tree construction algorithm which arranges all nodes in a way that each parent will have the maximal-available energy resources to receive data from all of its children. And the minimum energy node is used to collect data for data aggregation . Such arrangement will extend the time to refresh the tree and lowers the amount of data loss due to a broken tree link before the tree reconstructions. We proceed to construction of Centralized Lifetime Maximizing Tree structure .We assume that the complete knowledge of the event region including the connectivity and residual energy of all the source nodes is known prior to the start of this construction. The simple way to obtain CLMT is to directly run an extensive search at each node and then compare their tree energies. This method is very simple but has the scalability problem ,when the network starts to grow or becomes dense the search becomes more wide and takes a lot of time as well as the number of comparisons increase. So we tackle this issue with a complete different approach .

CLMT requires a root (initially unknown) to collect data from every other node via routes with the highest branch energy subject to condition that loop is not created. We therefore follow this convention, and define the branch energy as the minimum energy of all the non-leaf nodes in a given branch.

Let

$brE_{x,y,A}$ : Energy of branch $A_x$ leafed at node x and rooted at node y, $A \in P_{x,y}$.

Branch energy is calculated as

$$brE_{x,y,A} = \min_{i \in A_x, i \neq x} \{e_i\} \qquad (1)$$

Tree to be constructed should have energy that directly depends on the minimum residual energy of all non-leaf nodes. So our main aim is to find out this minimum-energy node .This node represents the bottleneck to the network. If this bottleneck node is detected, it will then be easy to determine what the highest energy tree will be . To illustrate above descriptions, consider the set of nodes shown in figure 1. Any source node can either be a root, parent or leaf. Figure 1(a) shows the connectivity diagram of set of nodes . By assuming that node **e** is a root, node **g** has to collect data from nodes **b** and **c** as shown in figure 1(b) because tree energy has to be the highest. If node **e** is now a parent ,node **g** again has to forward data for nodes **b** and **c** , in the network as shown in figure1(c). Finally if node **e** is a leaf , the protocol must have node **h** to forward its data to some root via either nodes **d** or **g** . By using some argument, node **g** has to be again parent for data collection from node **h** as shown in figure 1(d) .We therefore call node **g** as **bottleneck node** (focus node ) for this particular network scenario, since there are no better ways to route around this node. CLMT must be rooted at this node and the tree must have energy less than that of this node.

### A. Identification of Bottleneck Node

Our main aim is to identify this bottleneck node and coordinate the given set of network connection such that a tree is obtained with this node being configured as the minimum energy non-leaf node. For this issue we begin by arranging nodes in the ascending energy levels. Starting from the least-energy node ,we test if the removal of all network links to this node except from its highest-energy neighbor will disconnect the existing graph. If so ,bottleneck node is found and there are no better ways than to collect data via this node. The removed links are thus restored and any tree rooted at one of the nodes in the remaining set shall have the energy as that of this chosen node. If the above mentioned condition is not satisfied then removed links do not contribute to the construction of the CLMT and we shall move to the next node. The energy of the highest-energy neighbor has to be greater than that of the node under the test. When such neighbor does not exist, the node has to be a parent for at least one of its neighbors, and thus all the links are preserved and the lifetime is maximized. In the case when there are more than one neighbors that have equal highest energy, either one can serve as the parent for collecting data from the node under the test without affecting the tree energy. Node ID is thus used to break this tie. Finally, when we come to the last node, i.e. the highest-energy one, we conclude that there is no bottleneck node for this particular topology and any tree rooted at this last node, on the existing graph, can have the highest tree energy. This tree will be known as CLMT in this case .To illustrate the description , consider two examples. Figure 2 depicts the CLMT search during which a bottleneck node is found when links are removed. As shown in figure 2(a) starting from the least energy node i.e node **d** , we test if the removal of all the links to this node except that from its highest-energy neighbor will disconnect the existing graph. The process is proceeded on all nodes following the ascending energy level. The same test is done on node **f** as shown in figure 2(b) , the same way node **c** is tested in figure 2(c) .For this network when we test node **g** and the link from nodes **g** to **b** is removed the existing graph gets disconnected and node **g** has to be a parent for some nodes in the network. So node **g** is referred as the **bottleneck node**. Any tree rooted at node **g** will be the CLMT. Any tree rooted at one of the nodes in the remaining set i.e nodes **b, h, e, g, or a** therefore have the highest tree energy of **7 J** as that of this bottleneck node (shown in figure 2(d)).

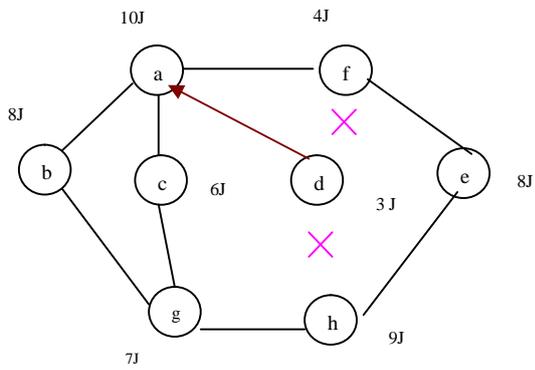
(a) Node **d** under test

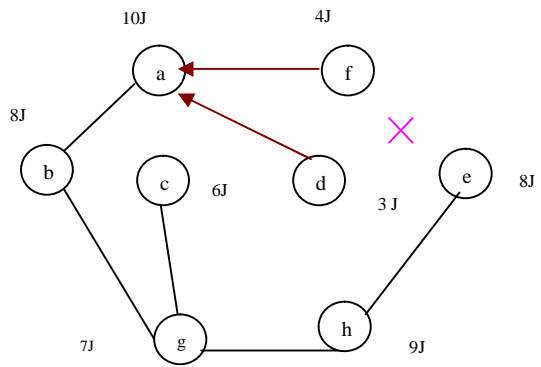
(b) Node **f** under test

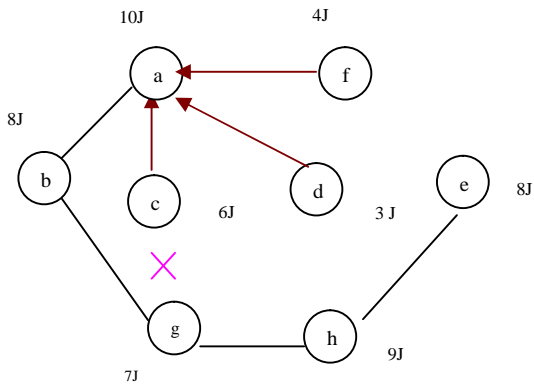
(c) Node **c** under test

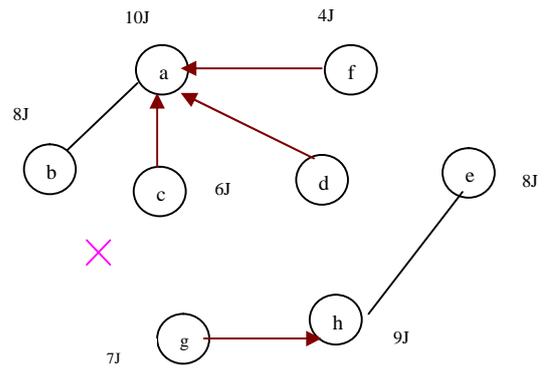
(d) Bottleneck node **g** found

**Figure 2 An example to search CLMT where bottleneck node is found**

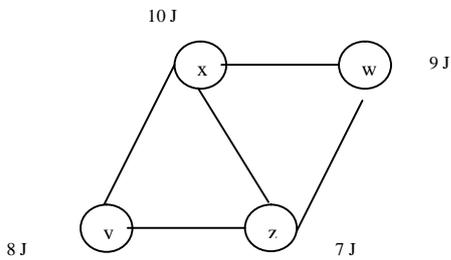
(a) 4- node topology

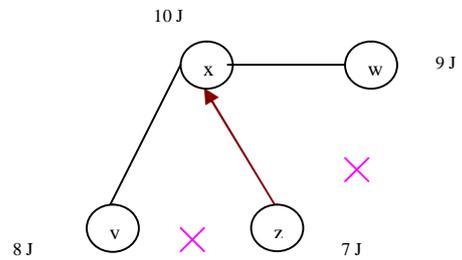
(b) Node **z** under test

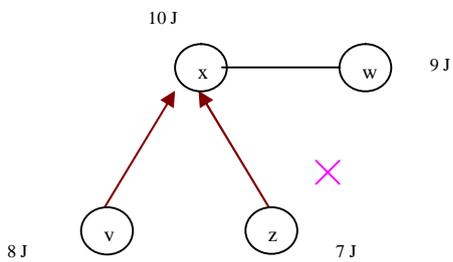
(c) Node **y** under test

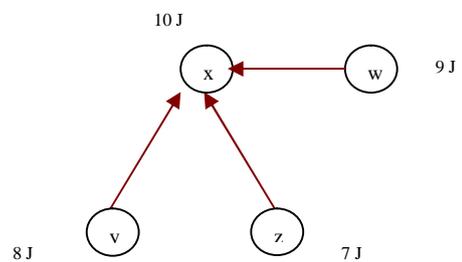
(d) bottleneck node not found

**Figure 3  An example to search CLMT where bottleneck node is not found**

Figure 3 depicts another example of the search of 4- node topology where no bottleneck node is found. We start testing from least-energy node and continue till last. Removal of the links ,expect that from the highest-energy neighbor ,to the source node under test does not disconnect the existing graph, hence bottleneck node is not found for this particular topology. Therefore any tree rooted at the highest-energy node, i.e node **x** will be the CLMT .Node **x** will be the node for data collection for this topology. The concept is shown in figure 3(a) - 3(d) .

*B. Proposed CLMT Algorithm*

---

***Define :***

$node_i$ to be the node with $i^{th}$ least energy
$e(node_i)$ to be the energy of $node_i$
$node_{i,max}$ to be the highest-energy neighbor of $node_i$
subject to the condition that $e(node_{i,max}) > e(node_i)$
$link_{a,b}$ to be the bi-directional link between nodes a and b
$TE_i$ to be the energy of a tree rooted at node i
N to be the set , initially empty

***Centralized Lifetime Maximizing Tree : Algorithm***

1) sort nodes in ascending energy level
2) for i = 1 to I , i ++
3) get $node_{i,max}$
4) if $node_{i,max}$ exists,
5) remove $link_{i,n}$ and store n in N ,
$\forall n \in I, n \neq node_{i,max}, n \neq i$
6) if the graph is not connected,
7) restore $link_{i,j}$ $\forall j \in N$ and clear N
8) set $node_k$ to be the root , run Dijkstra's algorithm on $node_k$ where k is any one number from i to I
9) set $TE_k$ to be $e(node_i)$
10) return
11) set $node_I$ to be the root and run Dijkstra's algorithm on $node_I$
12) compute tree energy $TE_I$ for the tree rooted at $node_I$

---

The proposed algorithm for CLMT is summarized above. This algorithm creates a centralized lifetime maximizing tree spanning all the source nodes as vertices and takes care that no loops are created. Line 1 sorts all nodes in ascending energy levels. Lines 2 and 3 compute the highest-energy neighbor for the node under the test. Recall that the energy of this neighbor has to be greater than that of the node. When such a neighbor exists, lines 4 and 5 remove and temporarily store all links to the node except that from the highest-energy neighbor. Lines 6 to 10 restore the removed links, clear the storage, and compute a tree by running Dijkstra's algorithm [14] at one of the nodes in the remaining set when a bottleneck node is found. Tree energy is set to the energy of the bottleneck node at this time. In fact, the reason to run the Dijkstra's algorithm is to ensure that the remaining set of network connections does not create loops. Lines 11 to 12 compute a tree again by running Dijkstra's algorithm at the highest-energy node, and search the tree energy by using Equation 2 when no bottleneck node exists in the network. Tree energy is defined as the minimum branch energy of all the branches in a given tree.

Let

$I_z$ : Set of nodes in given tree rooted at node z
$TE\ I_z$ : Energy of tree rooted at node z

Tree energy is calculated as :

$$TEI_z = \min_{j \in I_z, j \neq x} \{e_j\} \quad (2)$$

## IV. SUMMARY

To meet the demands of WSNs where raw data readings are usually aggregated along their ways to be gathered at single source prior to transmissions to any interested sink. We have proposed in this paper a novel Centralized Lifetime Maximizing Tree construction algorithm for future wireless sensor networks. This tree provides a given set of sources with a mechanism to collect their data so that only a minimum amount of energy is required to deliver the same amount of information to the sinks. We began with an investigation to the conventional spanning tree and the energy aware variant of it for their uses in data aggregation. The conventional spanning tree fails to consider residual energy of nodes in the tree constructions. There is thus a good possibility that a low-energy node is arranged to forward data for some other nodes. E-Span improves the design of tree construction by assigning root to be the highest energy node. Such arrangement provides root with the maximum amount of energy resources for its additional duty in coordinating the route to distant sinks. However, there is still a high chance of assigning low-energy nodes to be the data aggregating agents for the other sources. To shorten the time and minimize the energy cost and hence maximize the functional lifetime of all sources, we have proposed a Centralized Lifetime Maximizing Tree construction algorithm which identifies the node that is causing a bottleneck to the set of connectivity provided by various event sources. Such arrangement extends the time to refresh the tree and lowers the amount of data loss due to broken tree link before the tree reconstructions. As well as this tree minimizes the delay and maximizes the lifetime of the source events by using minimum energy node as data collection node. In future we will simulate and compare the proposed CLMT tree with Conventional spanning trees, E-Span and other existing tree structures.